# Cepstral Scanning Transmission Electron Microscopy Imaging of Severe Lattice Distortions

Yu-Tsun Shao[1,2*], Renliang Yuan[1,2*], Haw-Wen Hsiao[1,2*], Qun Yang[1,3], Yang Hu[1,2], Jian-Min Zuo[1,2&]

[1]Department of Materials Science and Engineering, University of Illinois at Urbana-Champaign, 1304 W Green St, Urbana, IL 61801, United States

[2]Fredrick Seitz Materials Research Laboratory, University of Illinois at Urbana-Champaign, 104 S Goodwin Ave, Urbana, IL 61801, United States

[3]School of Physical Science and Technology, ShanghaiTech University, Shanghai, China

*First three authors contributed equally to this work

Correspondence: jianzuo@illinois.edu

**Abstract**

The development of four-dimensional (4D) scanning transmission electron microscopy (STEM) using fast detectors has opened-up new avenues for addressing some of long-standing challenges in electron imaging. One of these challenges is how to image severely distorted crystal lattices, such as at a dislocation core. Here we introduce a new 4D-STEM technique, called Cepstral STEM, for imaging disordered crystals using electron diffuse scattering. Local fluctuations of diffuse scattering are captured by scanning electron nanodiffraction (SEND) using a coherent probe. The harmonic signals in electron diffuse scattering are detected through Cepstral analysis and used for imaging. By integrating Cepstral analysis with 4D-STEM, we demonstrate that information about the distortive part of electron scattering potential can be separated and imaged at nm spatial resolution. We apply our technique to the analysis of a dislocation core in SiGe and lattice distortions in high entropy alloy.

**Introduction**

The determination of severe lattice distortions around defects in real materials remains as a challenging problem in materials characterization. Without detailed experimental knowledge, our understanding of key materials properties, such as energy and mobility of dislocations and deep states responsible for recombination in semiconductors, has relied on first-principles calculations for relatively simple systems, such as metals and semiconductors and their alloys [1, 2]. In real materials, the interplay between composition, defect and strain makes such calculations based on theory alone intractable. The identification of these defects has largely relied on their extended strain field for diffraction contrast imaging. While the continuum elasticity theory suffices in modelling long-range elastic interactions , images obtained near dislocation core or from complex defects are difficult to interpret [3].

Atomic resolution imaging, in principles, would allow an observation of severe lattice distortions, such as dislocation kinks and jogs at defect core. Recent investigations of dislocation core structures by aberration corrected scanning transmission electron microscopy (STEM) [4] and high resolution electron microscopy (HREM) [5] have demonstrated such possibilities for the study of compound semiconductors and oxides, including complex oxides [6-10]. The benefits of having <1 Å resolution with aberration correction, compared to 2 Å resolution in conventional

TEM/STEM, are most dramatic in the observation of diamond or zinc blende structure along [110], where the projected atomic columns can be resolved. Most dislocation core structure observations were made edge-on with two notable exceptions, one is the imaging of screw dislocations in GaN [11] and the other is by Prof. Spence and his collaborators on imaging dislocation kinks in silicon [12, 13]. Both used the electron beam perpendicular to the dislocation line. In both cases, novel contrast mechanisms, using depth of focus and forbidden reflections, respectively, were employed for dislocation imaging.

Using computer simulations, Spence and Koch suggested coherent convergent beam diffraction (CBED) for the determination of dislocation core structure. The idea is to use an electron beam positioned along the dislocation line to record electron scattering between Bragg reflections by the dislocation for structure determination [14, 15]. Such experiment requires the positioning of the electron beam within 1 nm or better accuracy to the dislocation core, which was difficult at the time that this idea was proposed. This has changed with the development of scanning based diffraction techniques that enables the recording of 4D electron diffraction datasets (4D-EDDs), i.e., the electron wave intensity in the momentum space ($k_x$, $k_y$) for each probe position in the real space ($x$, $y$) and thus 4D [16]. The development of fast electron detectors has generated further excitements in using 4D-EDDs for STEM imaging or 4D-STEM [17, 18]. Here, we demonstrate a new form of 4D-STEM, called Cepstral STEM, to image severe lattice distortions. Cepstral STEM uses the real-space information provided by coherent electron nanodiffraction patterns recorded using a small convergence beam. A major advantage of Cepstral STEM is the separation of Bragg diffraction from electron diffuse scattering. The separation allows the use of electron diffuse scattering for the determination of the distortive part of electron scattering potential, which was not possible before in STEM.

## Methods and Theory

### Acquisition of 4D Electron diffraction datasets

Electron diffraction was carried using a probe $C_s$-corrected Themis Z STEM (Thermo Fisher Scientific, USA), operated at 300 kV. Experimental 4D-EDDs were acquired over regions of interest using an electron probe of ~1.1 mrad in semi-convergence angle (θ) and 1.2 nm in FWHM (full-width at half-maximum). The electron probe was formed in the µProbe mode of Themis Z with a Schottky field emission gun. A small condenser aperture (50 micron) was used for forming the probe, and electron illumination within the aperture can be considered as coherent [19].

The TEM samples of SiGe and HEA were prepared by the focused ion beam (FIB) method, initially using a 30 keV ion beam for cutting and thinning, and then a 5 keV beam for polishing to the targeted thicknesses. The dislocations in SiGe were also imaged at atomic resolution using HAADF-STEM. The STEM image was collected in Themis Z using a focused probe with a semi-convergence angle of 18 mrad and beam current of 30 pA, and an ADF detector of inner cutoff angle of 49 mrad.

Electron diffraction patterns were recorded using a scintillator coupled CMOS camera (Ceta camera, Thermo-Fisher) with 4096x4096 pixels, pixel size of 14 µm and a dynamic range of 16 bits. The center areas of 1024x1024 or 512x512 was used for recording with exposure times ranging from 20 ms to 0.1 s for each diffraction pattern, dependent on the strength of electron diffuse scattering.

**Cepstral analysis of electron nanodiffraction patterns**

Cepstral analysis is a sensitive signal processing technique for detecting weak harmonic signals [20, 21]. When applied to electron diffraction patterns, cepstral analysis is capable of measuring small amount of lattice strain [22]. A Cepstral pattern is obtained using

$$C_p = \left| FT\left\{ \log\left[ I(\vec{k}) \right] \right\} \right|. \quad (1)$$

Where $\vec{k}$ is the diffracted electron wave vector and $I$ for diffraction intensity. Figure 1a shows an example $C_p$ calculated from the nanodiffraction pattern taken from a silicon sample along [110] (Fig. 1b). Since diffraction pattern is in reciprocal space, the Cepstral peaks (Fig. 1b) detect harmonic signals in real space, or distances in Å, with zero distance at the center of the pattern. The Cepstral intensity decreases as the distances increase. The damping is caused by the electron probe shape, which limits the sample diffraction volume and thus the largest measurable interatomic distances. Figure 1c and d explain the distances recorded in $C_p$. The smaller distance between two silicon atoms in a dumbbell is not revolved here, since the nanodiffraction patterns recorded here are dynamical and the intensity of (002) is not zero because of electron multiple scattering. The dumbbells can be resolved for thin samples using kinematical CBED, for example [23].

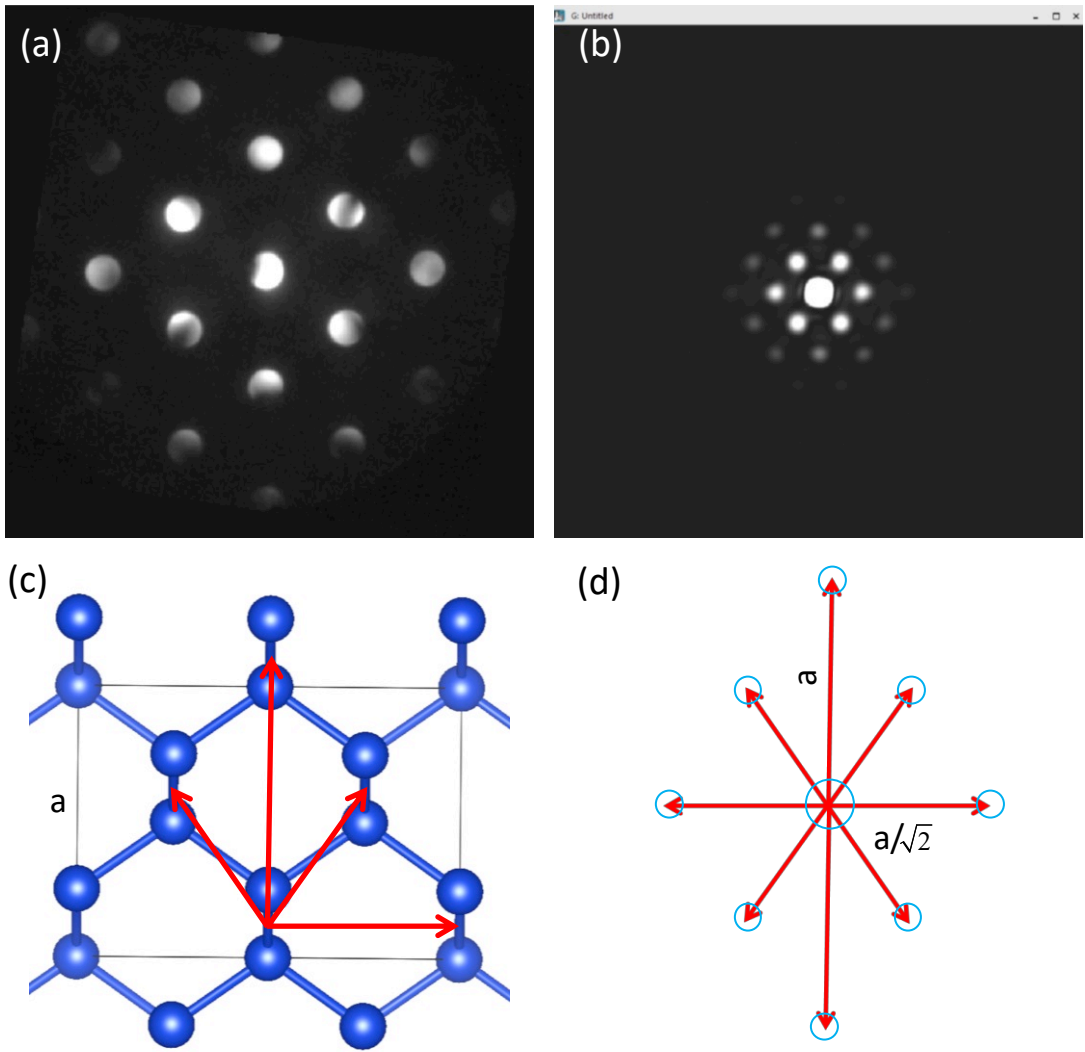

*Figure 1 Cepstral transform of electron nanodiffraction pattern. (a) Experimental diffraction pattern from a SiGe sample along [110], (b) Cepstral pattern from (a), (c) a model of silicon diamond structure projected along [110] with marked inter-dumbbell vectors, and (d) Inter-dumbbell vectors correspond to peaks in (b).*

To extend cepstral analysis for electron diffuse scattering analysis, we take advantage of the averaged and local structural information captured in a 4D-EDD, by calculating the difference between the Cepstral transforms of a local nanodiffraction pattern and the region averaged nanodiffraction pattern. In what follows, we show that the resulting differential Cepstrum ($dC_p$) approximately corresponds to the autocorrelation function, or Patterson function (PF) [24], of the distortive part of scattering potential in a thin sample. The $dC_p$ is calculated according to

$$dC_p = \left| FT \left\{ \log \left[ \frac{I(\vec{k})}{I_0(\vec{k})} \right] \right\} \right| = \left| FT \left\{ \log \left[ I(\vec{k}) \right] \right\} - FT \left\{ \log \left[ I_0(\vec{k}) \right] \right\} \right| \quad (2)$$

Where $I_0(\vec{k})$ represents intensity in the area averaged pattern, while $I(\vec{k})$ is the intensity in a single pattern from the 4D-EDD. The interpretation of $dC_p$ can be made based on the separation of the fluctuating part of the scattering potential ($U_1$) from the average scattering potential ($U_o$). The scattering potential seen by a small coherent electron probe is a sum of these two:

$$U = U_o + U_1, \quad (3)$$

where $U_1$ varies with the electron probe position. Diffraction by $U$ gives the diffraction pattern $I(\vec{k})$, while diffraction by $U_o$ gives $I_0(\vec{k})$. If we assume the fluctuations are random, then $\langle U_1 \rangle = 0$.

The distortive part of scattering potential is the origin of diffuse scattering, which is seen in between Bragg reflections. In electron diffraction, diffuse scattering is often observed using selected area electron diffraction [25, 26]. In a nanodiffraction pattern obtained using a coherent electron beam, the diffuse scattering is more like laser speckles, in a way similar to fluctuations recorded in amorphous materials [27]. The speckle pattern is averaged over the sample thickness. The diffraction pattern $I(\vec{k})$ thus has two parts under the approximations that electron diffuse scattering $I_D(\vec{k})$ is weak and the wave function associated with diffuse scattering has a random phase [19, 28]

$$I(\vec{k}) = I_o(\vec{k}) + \bar{I}_o(\vec{k}) * I_D(\vec{k}), \quad (4)$$

where $\bar{I}_o(\vec{k})$ is the thickness averaged diffraction intensity. The convolution and thickness averaging reflect that there are many beams in $I_0(\vec{k})$, each contributes to the total electron diffuse scattering through the entire sample thickness [19, 28]. Using

$$\log\left[I(\vec{k})\right] \approx \log I_o(\vec{k}) + \Lambda(\vec{k}) * I_D(\vec{k}) \qquad (5)$$

Where $\Lambda(\vec{k}) = \bar{I}_o(\vec{k}) / I_o(\vec{k})$. FT of $\Lambda(\vec{k}) * I_D(\vec{k})$ then gives

$$\left|FT\left[\Lambda(\vec{k})\right]FT\left[I_D(\vec{k})\right]\right| = \left|FT\left\{\log\left[\frac{I(\vec{k})}{I_o(\vec{k})}\right]\right\}\right| = dC_p \qquad (6)$$

Thus $dC_p$ gives the Patterson function of the fluctuating scattering potential multiplied by a shape function. This shape function is approximately the FT of a top-hat function in a diffraction pattern where the transmitted beam is much stronger than the diffracted beams.

The sensitivity of $dC_p$ to the distortive part of potential is demonstrated in Figure 2, where two examples are selected from a SiGe sample with Fig. 2a away from and Fig. 2c at a dislocation. The patterns are shown at the same intensity scale for comparison. Strong speckles are observed in Fig. 2c, while Fig. 2a is more like an electron diffraction pattern with diffuse scattering. The $dC_p$ of Fig. 2b is weak with no strong harmonic signals. Compared with Fig. 2b, Fig. 2d shows strong harmonic peaks in the direction normal to the streaky speckles observed in Fig. 2c. Thus, both the intensity and harmonic peaks are sensitive to the amount and type of electron diffuse scattering.

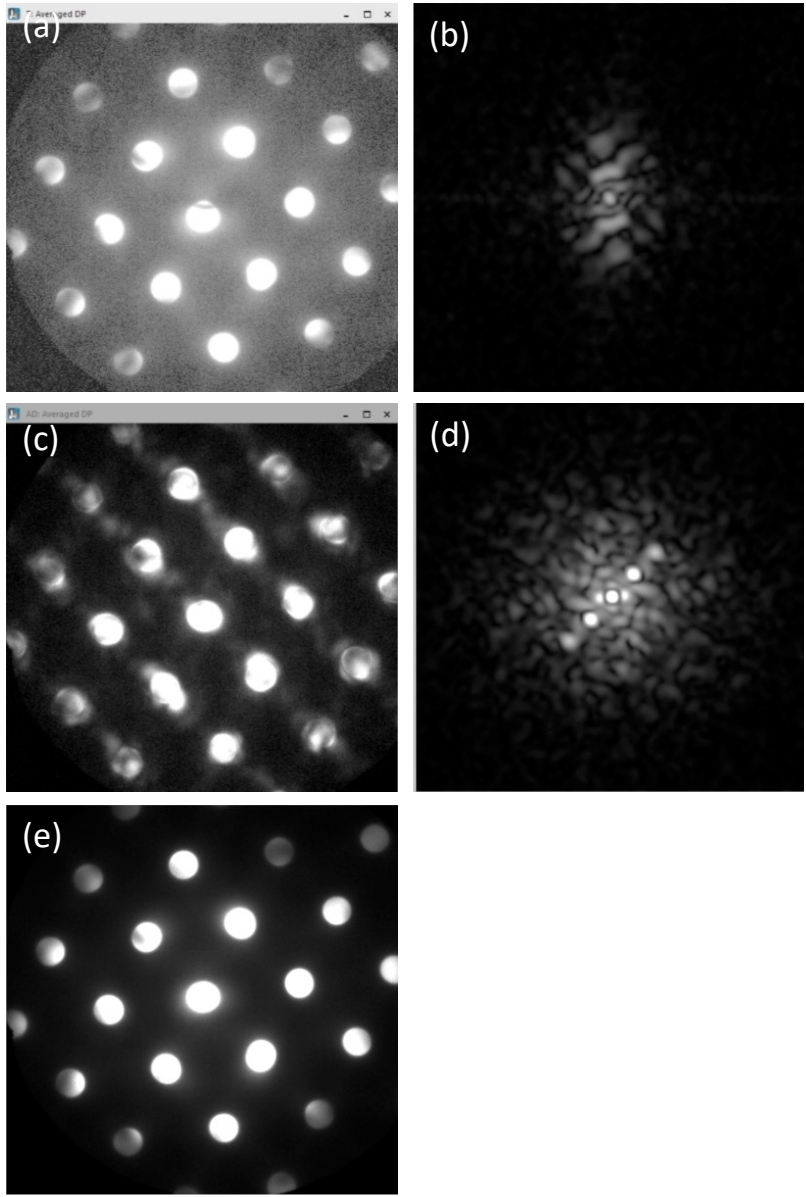

*Figure 2 Two examples of $dC_p$. (a) An electron nanodiffraction pattern away from dislocations in a SiGe sample along [110]. (b) $dC_p$ of (a). (c) An electron nanodiffraction pattern at a dislocation core in the SiGe sample, (d) $dC_p$ of (c). (e) The average pattern from a 40x40 4D-EDD, which is used for the calculation of $dC_p$. For comparison, (a), (c) and (e) are displayed at the scale of 0-500 a.u. and (b) and (d) at the scale of 0-6500 a.u..*

## Cepstral STEM

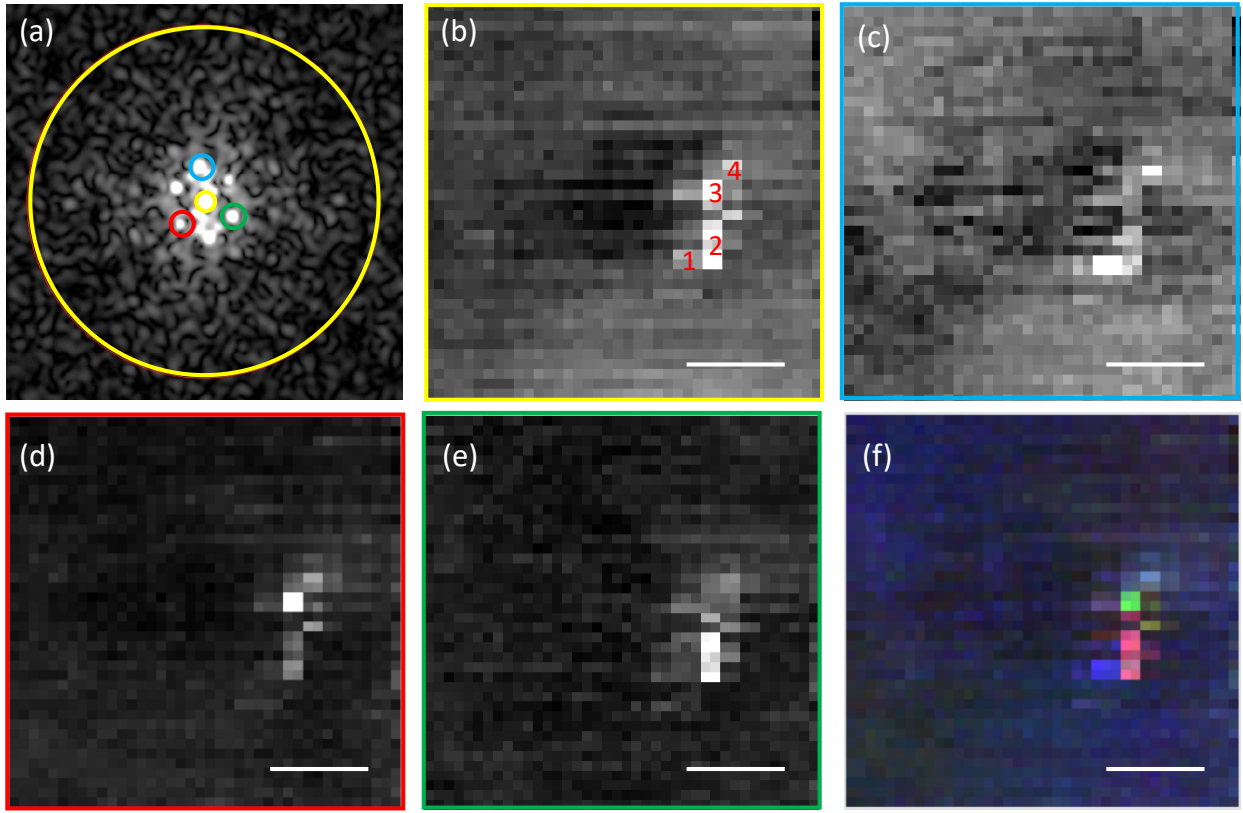

*Figure 3 Cepstral STEM imaging of dislocation core. (a) 4D Cepstral data where intensity within the marked circles is integrated to form Cepstral ADF image of (b) ADF between yellow circles. (c) DF within blue circle, (d) and (e) within the red and green circles, respectively. (f) The RGB image formed using (c), (d) and (e). The 4D-EDD was acquired from a SiGe sample with edge-on dislocations over the area of 40x40 $nm^2$. The scale bar is 10 nm.*

The advantage of having 4D-EDDs is that electron diffraction intensity can be analyzed and related to the structure of samples [29, 30] or the electric and magnetic fields for imaging [31, 32]. As the speed of detectors improves dramatically, 4D-EDDs can be collected over larger areas than previously possible [33], which makes 4D-STEM more attractive than direct STEM imaging for some applications [17]. The examples of structural analysis include orientation [34, 35], domain [36] and strain mapping [29]. In all these cases, the diffraction signals from Bragg diffraction are analyzed in reciprocal space and mapped in real space. The Cepstral analysis method for strain mapping introduced by Padgett et al. [22] measures distances directly in real space but again relies on Bragg diffraction. The idea of Cepstral STEM is thus to take advantage of electron diffuse scattering for imaging fluctuations in electron scattering potential.

Figure 3 demonstrates an example of Cepstral STEM imaging of edge-on dislocations in a SiGe sample, which was grown on top of silicon. Because of the lattice strain, the sample contains both misfit and threading dislocations [37]. A 4D-EDD was collected from a location with edge-on misfit dislocations. The scan is over an area of 40x40 nm$^2$ and the step size of 1 nm. Figure 3a represents a $dC_p$ pattern obtained from the 4D-EDD using the methods described in Fig. 3. A Cepstral ADF image (Fig. 3b) is obtained by integrating the $dC_p$ intensity between the two marked circles in Fig. 3a. The same principle can also be used to form bright or dark field images. Figure 3c, d and e show three Cepstral dark field images obtained by integrating three different Cepstral peaks marked in Fig. 3a. The contrast in Fig. 3b represents the magnitude of distortive potential, which shows high contrast at the dislocation core region. Figures 3c, d and e give different contrast, the strong contrast in each figure is associated with the regions where a particular Cepstral peak is strong. Putting them together, the composite image of Fig. 3f demonstrates the magnitude as well as the harmonics in of distortive potential at the dislocation core.

 Applications

**Cepstral STEM imaging of dislocation core in SiGe**

Here we demonstrate the principle using Cepstral STEM for imaging the core structure of the misfit dislocations observed in Fig. 3. To help with the interpretation of the obtained $dC_p$ harmonic signals, we also acquired an atomic resolution HAADF-STEM image from the region studied by SEND. The image reveals a stacking fault bounded by two 30º partial dislocations with Burges vectors of $b_1 = 1/6[\bar{2}\bar{1}\bar{1}]$ and $b_2 = 1/6[211]$. A perfect dislocation b3 with a Burgers vector of $1/2[\bar{1}0\bar{1}]$ is found near b2. The b2 and b3 are separated by 3 atomic layers.

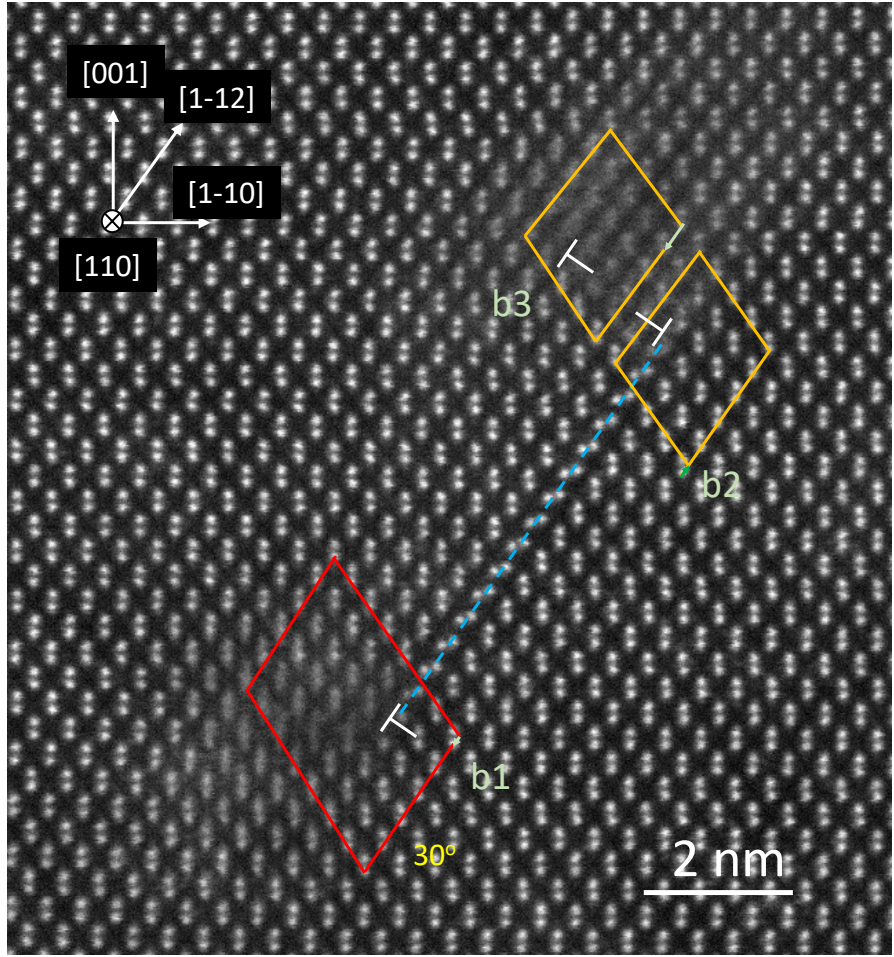

*Figure* 4 *Atomic resolution HAADF-STEM image of dislocations associated with a stacking fault.*

Figure 5 shows the distinctive $dC_p$ patterns obtained from four areas of the dislocation core in SiGe, as marked in Fig. 3b. The areas are approximately located in the atomic resolution image of Fig. 4 and selected for comparison here. In Area 2, the $dC_p$ peaks correspond to the inter-dumbbell distances along [1-12], where the change in the dumbbell direction in the stacking fault introduces a one-dimensional distortive potential in 2D projection. Interestingly, the $dC_p$ pattern is insensitive to the relative shift between the two lattices across the stacking fault. This insensitivity can be attributed to the lack of interference signals in the recorded diffraction patterns, which presumably is weak. In Area 3, the $dC_p$ peaks can be associated with the edge dislocation with the missing half plane along [-112] and distortions around the dislocation as marked by yellow and red lines, respectively. The $dC_p$ pattern in Area 4 is characterized by the reduced second order peaks along [-112] and enhanced peaks along [1-10]. The strong first order peaks indicate distortions that are not immediately obvious in the atomic resolution image as in Areas 2 and 3. In Area 1, two first order peaks are observed in the $dC_p$ pattern along [1-10], which can be attributed to atomic distortions behind the smearing of dumbbell contrast along the line indicated by the arrow. These results indicate the sensitivity of Cepstral STEM to atomic distortions in the dislocation core.

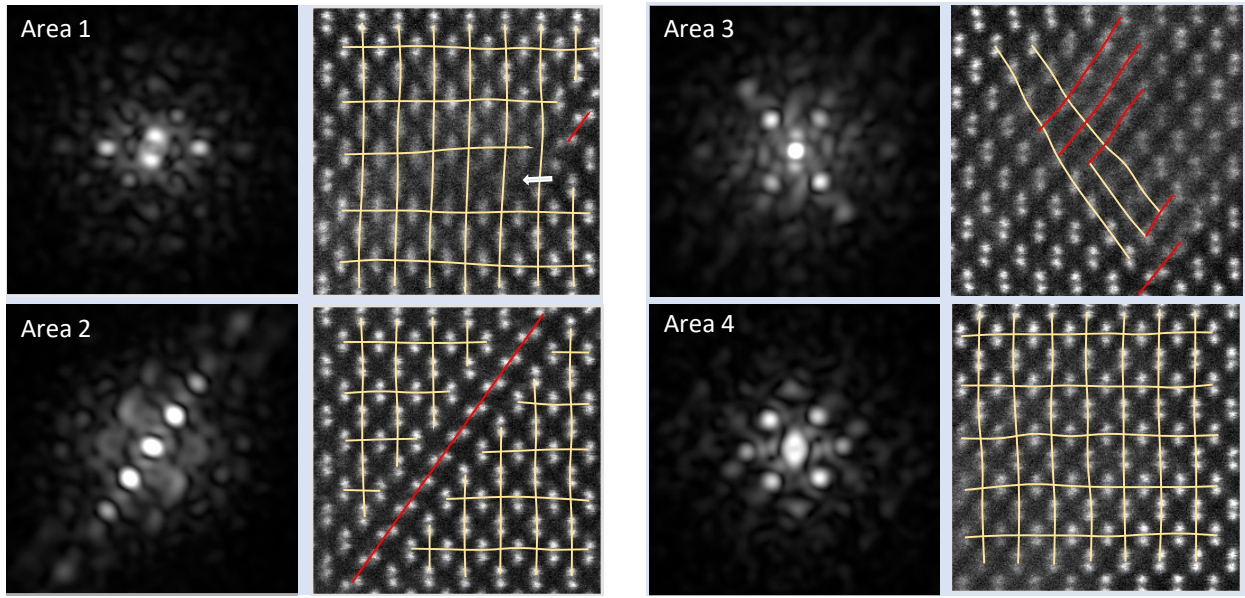

Figure 5 Differential Cepstral patterns from 4 areas of dislocation core in SiGe and comparison with corresponding atomic resolution HAADF-STEM images.

A major utility of Cepstral STEM is to identify and extract electron nanodiffraction patterns from the severely distorted region from the 4D-EDD. Figure 6 shows the diffraction patterns from four areas identified in Figure 3b and Figure 5, where strong diffraction streaks in Figure 6 come from the stacking fault. Such diffraction patterns, in the future, can be combined with electron images to extract quantitative structural information about the dislocation core [38].

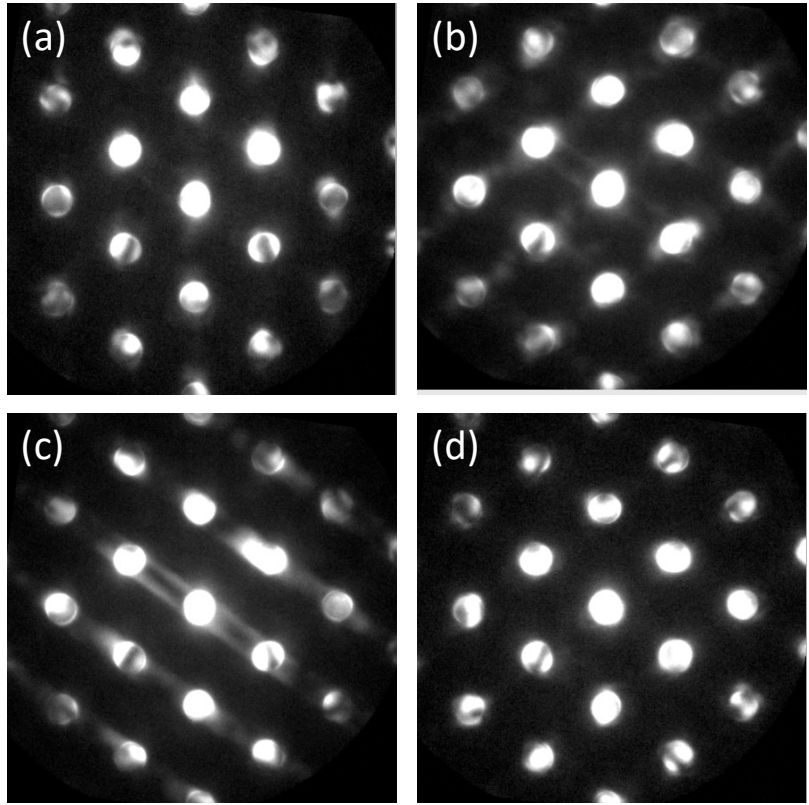

*Figure 6 Electron nanodiffraction patterns obtained from (a) Area 1, (b) Area 3, (c) Area 2 and (d) Area 4 of Figure 3b.*

**Imaging severe lattice distortion in high entropy alloy**

The above example demonstrated the ability of Cepstral STEM to detect harmonic signals in the distortive part of electron scattering potential and use these signals for imaging. Such ability is very useful for the study of a class of materials called strain glass that exhibit glass-like properties, such as relaxor ferroelectrics, pre-martenstic steel and complex alloys [39], where lattice distortion plays a critical role in materials' properties. Previous high-resolution characterization relied mainly on HREM and strain analysis using methods such as geometric phase analysis [40]. Other techniques for investigating such materials include symmetry mapping using CBED [41] and aberration-corrected STEM imaging [42], both have revealed local lattice distortions. To demonstrate the utility of Cepstral STEM for the characterization of complex lattice distortions, in what follows, we apply our new imaging technique to the study of high entropy alloy (HEA).

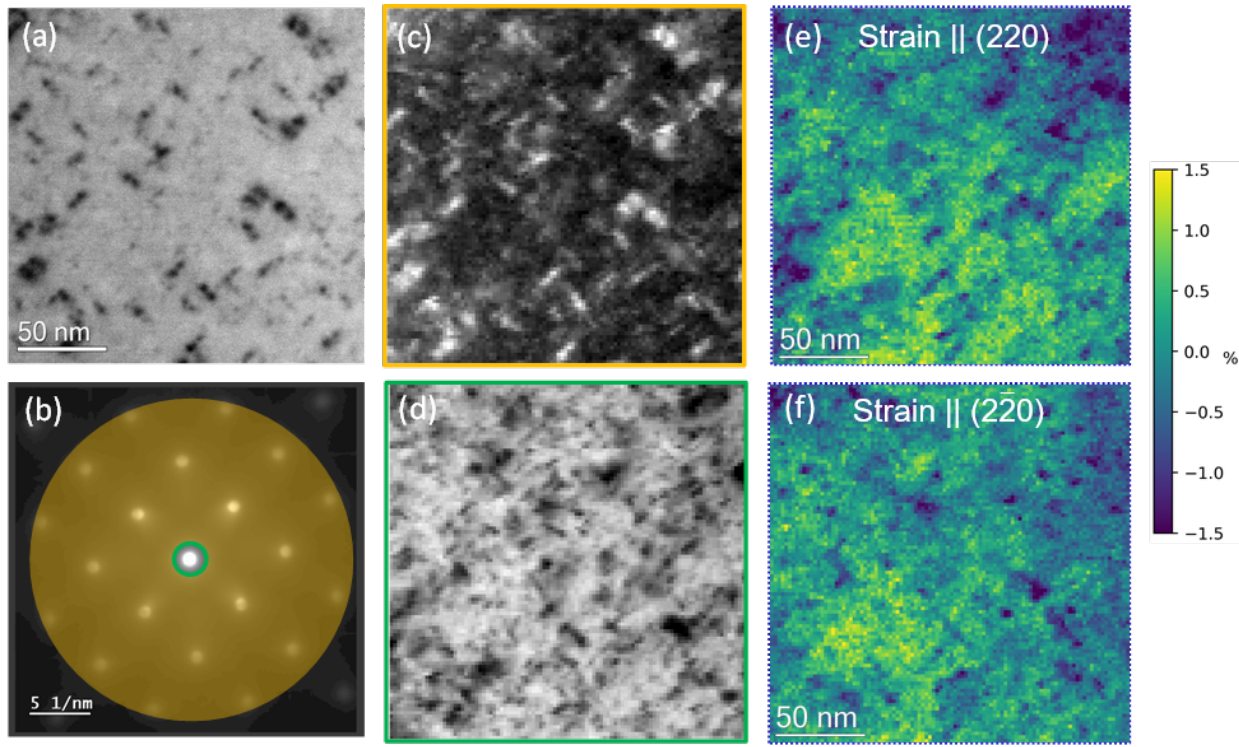

Figure 7 4D-STEM imaging of $Al_{0.1}CrFeCoNi$ based on a 4D-EDD acquired from the area in (a) using a 100x100 scan and a 2 nm step size. Three types of images are shown here using intensity integration for (c) ADF, (d) BF, and (e and f) strain maps.

The material we study here is $Al_{0.1}CrFeCoNi$, with a targeted composition of 2.44 at% for Al and 24.4 at% for Co, Cr, Fe, and Ni. The details about the preparation and structure characterization of $Al_{0.1}CrFeCoNi$, including anomalous diffraction contrast, can be found in Ref. [43]. This sample is selected here because HEAs are comprised of at least five elements in an equal or near-equal atomic fraction, and the mismatch in the atomic radii of different elements results in a large number of atoms being displaced from their ideal positions (lattice distortion). Consequently, severe lattice distortion has been suggested as one of the four core effects in HEAs

[44, 45]. However, the complex chemical and lattice disorder in the HEAs present a major challenge for their characterization [45]. For example, PDF analysis of atomic distances in CrMnFeCoNi and Ni alloys demonstrated no significant differences between the HEA and the conventional alloys [46].

Figure 7 shows an $Al_{0.1}CrFeCoNi$ HEA sample as observed by HAADF-STEM (Fig. 7a) and 4D-STEM (Figs. 7b to d). A 4D-EDD of 100x100 scan was acquired (Fig. 7b) using an electron nanoprobe scanning over the sample area of 200x200 $nm^2$ as shown in Fig. 7a. The total acquisition time is ~1 hr and during this period the sample drift was less than 3 nm. Figure 7c and d are virtual STEM images obtained from the 4D-EDD using the intensity integration masks shown in Fig. 7b for the BF and ADF STEM contrast. The ADF image gives the contrast complementary to the HAADF contrast in Fig. 7a. This result suggests that the dark contrast in Fig. 7a comes from diffraction effects, rather than compositional differences due to the Z-dependence of HAADF-STEM. The diffraction effects are further evidenced in the contrast obtained in the BF image (Fig. 7d). In Figs. 7e and f, we plot the strain maps obtained from the same 4D-EDD. The diffraction data was analyzed using the circular Hough transformation method described by Yuan et al. [29]. This method detects the diffracted disk position in a nanodiffraction pattern with high precision, using the filtered disk edge to reduce the impact of diffraction intensity. Inhomogeneous strain from -1.5% to 1.5% is observed, where the dark regions in the HAADF image (Fig. 7a) appear to be correlated with compressive strain. Together, Figure 7 demonstrates the types of information that one can extract based on diffraction intensity and geometry.

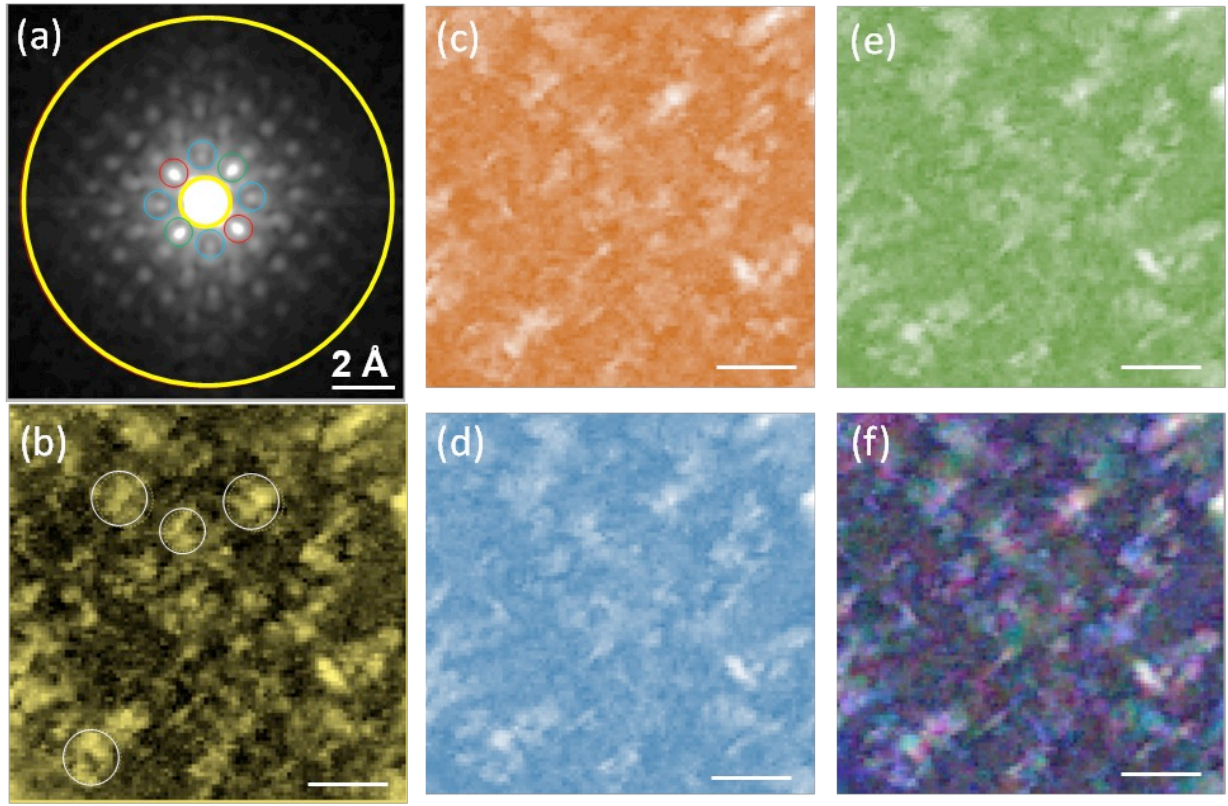

*Figure 8 Cepstral STEM of Al$_{0.1}$CrFeCoNi using the 4D-EDD acquired in Figure 7. (a) Areas averaged Cepstrum. (b) Cepstral ADF image obtained by integration between the marked yellow circles in (a). The white circles mark the areas where Cepstrum in (a) was obtained. (c), (e),(d) DF Cepstral STEM using intensities inside red, green and blue circles in (a), respectively. (f) Composite of (c), (d) and (e), for R, G and B colors. Scale bars in (b-f), 20 nm.*

Figure 8 shows the results of Cepstral STEM for the 4D-EDD described in Figure 7. The magnitude of distortive potential is imaged in Figure 7b, where bright contrast appears in the regions where strong diffraction contrast is observed in Figure 7c. The average of Cepstrum within the marked circles in Fig. 8b is shown in Fig. 8a, where strong first and second order and weaker high order harmonics are detected at 2.46 Å and 3.156 Å, respectively. The harmonic peak positions differ from those of a [001] projected fcc unit cell with interatomic distances of 1.795 Å or 2.539 Å, indicating the sensitivity of differential Cepstrum to distorted crystal potentials. Using the first order and second order harmonics, we formed DF Cepstral STEM images in Figs. 8c, d and e. The bright contrast in each of these images indicate strong signals for the selected harmonics.

Putting these images together into an RGB image, Fig. 8f shows complex distortional patterns in the HEA. The distortional pattern in Fig. 8f complements the strain maps shown in Figs. 7e and f, which provide information about the averaged crystal lattice, while Fig. 8 images the lattice distortions.

**Summary and Outlook**

We have introduced the principles using electron diffuse scattering for STEM imaging. The idea is based on the calculation of Cepstral difference between a local diffraction pattern and the average diffraction pattern in a four-dimensional electron diffraction dataset. We show that the Cepstral difference can be related to the Patterson function of diffuse scattering, arising from the distorted part of scattering potential. Using examples of dislocations in SiGe and high entropy alloy, we demonstrate the ways of harmonic signals detected by Cepstral analysis used to form Cepstral STEM images. The Cepstral STEM image contrast is compared with that of regular STEM and Bragg reflection based strain mapping. The results show that these image modes are complementary, with Cepstral STEM providing the critical information about severely distorted regions of crystals.

A critical component of Cepstral STEM imaging is coherent electron nanodiffaction and the capturing of fluctuations in electron diffuse scattering. Such fluctuations are averaged out in the traditional study of diffuse scattering. The interpretation of volume averaged diffuse scattering is difficult and often requires the help of modeling. Being able to directly image fluctuations in diffuse scattering at nm resolution and detect harmonic signals thus provides a way forward for studying the origin of diffuse scattering.

Looking forward, Cepstral STEM has the potential to transform how we study disordered crystals. As a complementary imaging method, it can be a useful tool in a range of applications, such as the study of defects in semiconductors and quantum materials, and in complex alloys for distortion enhanced properties. The sharp harmonic signals detected by Cepstral STEM also make numerical quantification of lattice distortion and their mapping possible. Future developments along this direction are expected to further the progress of developing electron nanodiffraction as a truly quantitative method.

**Acknowledgments:** The authors express many thanks to Profs. Peter Liaw for discussions and for providing the HEA sample. The experimental work on HEA was supported by DOE BES (Grant No. DEFG02-01ER45923). The study of SiGe was supported by Intel (Award #54071821). Q.Y. is additionally supported by the Centre for High-resolution Electron Microscopy (CħEM) of SPST at ShanghaiTech University under the Grant No. EM02161943. Electron microscopy experiments were carried out at the Center for Microanalysis of Materials at the Frederick Seitz Materials Research Laboratory of University of Illinois at Urbana-Champaign.

**References**

[1] J.C.H. Spence, Chapter 77 Experimental studies of dislocation core defects, Dislocations in Solids, 13 (2007) 419-452.
[2] D. Rodney, L. Ventelon, E. Clouet, L. Pizzagalli, F. Willaime, Ab initio modeling of dislocation core properties in metals and semiconductors, Acta Materialia, 124 (2017) 633-659.
[3] Z. Zhou, M.L. Jenkins, S.L. Dudarev, A.P. Sutton, M.A. Kirk, Simulations of weak-beam diffraction contrast images of dislocation loops by the many-beam Howie-Basinski equations, Philos Mag, 86 (2006) 4851-4881.
[4] S. Pennycook, P. Nellist, Scanning Transmission Electron Microscopy, Imaging and Analysis, in: Scanning Transmission Electron Microscopy, Imaging and Analysis, Springer, New York, 2011.
[5] J.C.H. Spence, High resolution electron microscopy, 4th Edition ed., Oxford University Press, Oxford, UK, 2013.
[6] Z.L. Zhang, W. Sigle, F. Phillipp, M. Ruhle, Direct atom-resolved imaging of oxides and their grain boundaries, Science, 302 (2003) 846-849.
[7] Z.L. Zhang, W. Sigle, M. Ruhle, Atomic and electronic characterization of the a 100 dislocation core in SrTiO3, Physical Review B, 66 (2002) 094108.
[8] C.L. Jia, M. Lentzen, K. Urban, Atomic-resolution imaging of oxygen in perovskite ceramics, Science, 299 (2003) 870-873.
[9] T. Paulauskas, C. Buurma, E. Colegrove, B. Stafford, Z. Guo, M.K.Y. Chan, C. Sun, M.J. Kim, S. Sivananthan, R.F. Klie, Atomic scale study of polar Lomer-Cottrell and Hirth lock dislocation cores in CdTe, Acta Crystallographica Section A, 70 (2014) 524-531.
[10] P.E. Batson, M.J. Lagos, Characterization of misfit dislocations in Si quantum well structures enabled by STEM based aberration correction, Ultramicroscopy, 180 (2017) 34-40.
[11] H. Yang, J.G. Lozano, T.J. Pennycook, L. Jones, P.B. Hirsch, P.D. Nellist, Imaging screw dislocations at atomic resolution by aberration-corrected electron optical sectioning, Nature Communications, 6 (2015) 7266.

[12] H. Alexander, J.C.H. Spence, D. Shindo, H. Gottschalk, N. Long, Forbidden-reflection lattice imaging for the determination of kink densities on partial dislocations, Philos Mag A, 53 (1986) 627-643.
[13] H.R. Kolar, J.C.H. Spence, H. Alexander, Observation of moving dislocation kinks and unpinning, Physical Review Letters, 77 (1996) 4031-4034.
[14] C. Koch, J.C.H. Spence, C. Zorman, M. Mehregany, J. Chung, Journal of Physics: Condensed Matter, 12 (2000) 10175-10183.
[15] J.C.H. Spence⊥, H.R. Kolar, G. Hembree, C.J. Humphreys, J. Barnard, R. Datta, C. Koch, F.M. Ross, J.F. Justo, Imaging dislocation cores – the way forward, Philos Mag, 86 (2006) 4781-4796.
[16] J.-M. Zuo, Electron Nanodiffraction, in: P.W. Hawkes, J.C.H. Spence (Eds.) Springer Handbook of Microscopy, Springer International Publishing, Cham, 2019, pp. 2-2.
[17] C. Ophus, Four-Dimensional Scanning Transmission Electron Microscopy (4D-STEM): From Scanning Nanodiffraction to Ptychography and Beyond, Microscopy and Microanalysis, 25 (2019) 563-582.
[18] Y. Jiang, Z. Chen, Y. Han, P. Deb, H. Gao, S. Xie, P. Purohit, M.W. Tate, J. Park, S.M. Gruner, V. Elser, D.A. Muller, Electron ptychography of 2D materials to deep sub-ångström resolution, Nature, 559 (2018) 343-349.
[19] J.M. Zuo, J.C.H. Spence, Advanced Transmission Electron Microscopy, Imaging and Diffraction in Nanoscience, Springer, New York, 2017.
[20] A.M. Noll, Cepstrum Pitch Determination, The Journal of the Acoustical Society of America, 41 (1967) 293-309.
[21] A.V. Oppenheim, R.W. Schafer, From frequency to quefrency: a history of the cepstrum, IEEE Signal Processing Magazine, 21 (2004) 95-106.
[22] E. Padgett, M.E. Holtz, P. Cueva, Y.-T. Shao, E. Langenberg, D.G. Schlom, D.A. Muller, The exit-wave power-cepstrum transform for scanning nanobeam electron diffraction: robust strain mapping at subnanometer resolution and subpicometer precision, Ultramicroscopy, 214 (2020) 112994.
[23] J.T. McKeown, J.C.H. Spence, The kinematic convergent-beam electron diffraction method for nanocrystal structure determination, Journal of Applied Physics, 106 (2009) 074309.
[24] A.L. Patterson, A Fourier Series Method for the Determination of the Components of Interatomic Distances in Crystals, Physical Review, 46 (1934) 372-376.
[25] J.M. Zuo, J. Pacaud, R. Hoier, J.C.H. Spence, Experimental measurement of electron diffuse scattering in magnetite using energy-filter and imaging plates, Micron, 31 (2000) 527-532.
[26] T.R. Welberry, Diffuse Scattering and Models of Disorder, Int. Union of Crystallography, Oxford University Press, 2010.
[27] P. Voyles, J. Hwang, Fluctuation Electron Microscopy, in: Characterization of Materials, John Wiley & Sons, Inc., 2002.
[28] J. Gjonnes, D. Watanabe, Dynamical diffuse scattering from magnesium oxide single crystals, Acta Crystallographica, 21 (1966) 297-302.
[29] R. Yuan, J. Zhang, J.-M. Zuo, Lattice strain mapping using circular Hough transform for electron diffraction disk detection, Ultramicroscopy, 207 (2019) 112837.
[30] Y.T. Shao, J.M. Zuo, Lattice-Rotation Vortex at the Charged Monoclinic Domain Boundary in a Relaxor Ferroelectric Crystal, Physical Review Letters, 118 (2017) 157601.

[31] J.A. Hachtel, J.C. Idrobo, M. Chi, Sub-Ångstrom electric field measurements on a universal detector in a scanning transmission electron microscope, Advanced Structural and Chemical Imaging, 4 (2018) 10.

[32] W. Gao, C. Addiego, H. Wang, X. Yan, Y. Hou, D. Ji, C. Heikes, Y. Zhang, L. Li, H. Huyan, T. Blum, T. Aoki, Y. Nie, D.G. Schlom, R. Wu, X. Pan, Real-space charge-density imaging with sub-ångström resolution by four-dimensional electron microscopy, Nature, 575 (2019) 480-484.

[33] J. Tao, D. Niebieskikwiat, M. Varela, W. Luo, M.A. Schofield, Y. Zhu, M.B. Salamon, J.M. Zuo, S.T. Pantelides, S.J. Pennycook, Direct Imaging of Nanoscale Phase Separation in $La_{0.55}Ca_{0.45}MnO_3$: Relationship to Colossal Magnetoresistance, Physical Review Letters, 103 (2009) 097202.

[34] E. Rauch, M. Véron, Improving angular resolution of the crystal orientation determined with spot diffraction patterns, Microscopy and Microanalysis, 16 (2010) 770-771.

[35] O. Panova, C. Ophus, C.J. Takacs, K.C. Bustillo, L. Balhorn, A. Salleo, N. Balsara, A.M. Minor, Diffraction imaging of nanocrystalline structures in organic semiconductor molecular thin films, Nature Materials, 18 (2019) 860-865.

[36] D.N. Johnstone, F.C.N. Firth, C.P. Grey, P.A. Midgley, M.J. Cliffe, S.M. Collins, Direct Imaging of Correlated Defect Nanodomains in a Metal–Organic Framework, Journal of the American Chemical Society, 142 (2020) 13081-13089.

[37] X. Chen, D. Zuo, S. Kim, J. Mabon, M. Sardela, J. Wen, J.-M. Zuo, Large Area and Depth-Profiling Dislocation Imaging and Strain Analysis in Si/SiGe/Si Heterostructures, Microscopy and Microanalysis, 20 (2014) 1521-1527.

[38] J.M. Zuo, J. Zhang, W.J. Huang, K. Ran, B. Jiang, Combining Real and Reciprocal Space Information for Aberration Free Coherent Electron Diffractive Imaging Ultramicroscopy, 111 (2011) 817-823.

[39] X. Ren, Y. Wang, Y. Zhou, Z. Zhang, D. Wang, G. Fan, K. Otsuka, T. Suzuki, Y. Ji, J. Zhang, Y. Tian, S. Hou, X. Ding, Strain glass in ferroelastic systems: Premartensitic tweed versus strain glass, Philos Mag, 90 (2010) 141-157.

[40] D. Schryvers, S. Van Aert, High-Resolution Visualization Techniques: Structural Aspects, in: T. Kakeshita, T. Fukuda, A. Saxena, A. Planes (Eds.) Disorder and Strain-Induced Complexity in Functional Materials, Springer Berlin Heidelberg, Berlin, Heidelberg, 2012, pp. 135-149.

[41] K.H. Kim, D.A. Payne, J.M. Zuo, Determination of fluctuations in local symmetry and measurement by convergent beam electron diffraction: applications to a relaxor-based ferroelectric crystal after thermal annealing, Journal of Applied Crystallography, 46 (2013) 1331-1337.

[42] A. Kumar, J.N. Baker, P.C. Bowes, M.J. Cabral, S. Zhang, E.C. Dickey, D.L. Irving, J.M. LeBeau, Atomic-resolution electron microscopy of nanoscale local structure in lead-based relaxor ferroelectrics, Nature Materials, (2020).

[43] Y. Hu, L. Shu, Q. Yang, W. Guo, P.K. Liaw, K.A. Dahmen, J.-M. Zuo, Dislocation avalanche mechanism in slowly compressed high entropy alloy nanopillars, Communications Physics, 1 (2018) 61.

[44] M.H. Tsai, J.W. Yeh, High-Entropy Alloys: A Critical Review, Materials Research Letters, 2 (2014) 107-123.

[45] D.B. Miracle, O.N. Senkov, A critical review of high entropy alloys and related concepts, Acta Materialia, 122 (2017) 448-511.

[46] L.R. Owen, E.J. Pickering, H.Y. Playford, H.J. Stone, M.G. Tucker, N.G. Jones, An assessment of the lattice strain in the CrMnFeCoNi high-entropy alloy, Acta Mater, 122 (2017) 11-18.